\documentclass[11pt,a4paper]{article}
\usepackage{jcappub}

\newcommand{\be}{\begin{equation}}
\newcommand{\ee}{\end{equation}}
\newcommand{\bea}{\begin{eqnarray}}
\newcommand{\eea}{\end{eqnarray}}

\title{Extending the generalized Chaplygin gas model by using geometrothermodynamics}

\author[a,b]{Alejandro Aviles,}
\author[a]{Aztl\'an Bastarrachea-Almodovar,}
\author[a]{Lorena Campuzano,}
\author[a,c]{and Hernando Quevedo}
\affiliation[a]{Instituto de Ciencias Nucleares, Universidad Nacional Autonoma de M\'exico, Mexico.}
\affiliation[b]{Departamento de F\'\i sica, Instituto Nacional de Investigaciones Nucleares, Mexico}
\affiliation[c]{Dipartimento di Fisica and ICRA, Universit\`a di Roma "La Sapienza", I-00185 Roma, Italy}

\emailAdd{aviles@ciencias.unam.mx}
\emailAdd{aztlanalmodovar@comunidad.unam.mx}
\emailAdd{loduque@gmail.com}
\emailAdd{quevedo@nucleares.unam.mx}

\abstract{We use the formalism of geometrothermodynamics (GTD) to derive fundamental thermodynamic equations that are used to 
construct general relativistic cosmological models. In particular, we show that the simplest possible fundamental equation, 
which corresponds in GTD to a system with no internal thermodynamic interaction, describes the different fluids of  the standard model of cosmology. 
In addition, a particular fundamental equation with internal thermodynamic interaction is shown to generate a new cosmological model
that correctly describes the dark sector of the Universe and contains as a special case the generalized Chaplygin gas model. 
}


\begin{document}
\maketitle


\section{Introduction}
\label{sec:int}

Geometrothermodynamics (GTD) is a formalism that has been developed during the past few years to describe ordinary 
thermodynamics by using differential geometry \cite{Quevedo:2006xk}. To this end, the states of thermodynamic equilibrium 
are considered as points of an abstract space called the equilibrium space ${\cal E}$. Furthermore, we associate to ${\cal E}$
a Riemannian metric $g$ in which all the geometric properties of ${\cal E}$ are encoded. In classical thermodynamics, all the properties
of a system can be derived from the fundamental equation \cite{Callen:Thermodynamics}; analogously, it can be shown that in GTD the explicit form of 
the metric $g$ can be derived from the fundamental equation. It is then expected that the thermodynamic properties of the system can be 
represented in terms of the geometric properties of ${\cal E}$. In particular, the curvature of  ${\cal E}$ could be associated with the
internal mechanical interaction between the constituents of the thermodynamic system, i.e. the thermodynamic interaction, so that curvature singularities, 
in turn, correspond to phase transitions.

In the above approach, starting from a particular fundamental equation, GTD provides the geometric structure of the corresponding equilibrium space. However, 
the formalism can also be used to generate fundamental equations. Indeed, if the metric $g$ is assumed to define an extremal surface embedded in a phase space ${\cal T}$ 
(this will be explained in detail in Sec. \ref{sec:gtd}), certain differential equations must be satisfied whose solutions turn out to be mathematically 
well-defined fundamental equations. The consequent question is whether this method can be used to generate fundamental equations that could be applied
to describe physical systems.
The main goal of this work is to show two particular cases that can 
be used to construct cosmological models in the framework of general relativity. The idea is to derive all the thermodynamic properties from the fundamental 
equations, and to use them as input to construct cosmological models.

Nowadays, the standard paradigm in the late-time description of the Universe is that it is homogeneous and isotropic when averaging over large scales, 
and that today it is dominated by two unknown forms of energy: dark energy, which accelerates the Universe, and dark matter that clusters by 
gravitational instability and is responsible for the formation of the structures  we see at a very wide range of scales in the cosmos. For a review on the current 
status of cosmology see \cite{CervantesCota:2011pn}. 

Because of the lack of a fundamental description of these two ingredients, several alternative proposals have appeared in the literature. 
In fact, the split of the dark sector into dark energy and dark matter is arbitrary, because  what we measure in gravitational experiments is the 
energy-momentum tensor of the total dark sector, a property that has been called {\it dark degeneracy} by M. Kunz in \cite{Kunz:2007rk}; see also 
\cite{Hu:1998tj,Rubano:2002sx,Wasserman:2002gb,Kunz:2009yx,Aviles:2010ui,Reyes:2011ij,Aviles:2011ak}. In part for this reason, over the last decade the models of 
unified descriptions of the dark sector have played an increasingly important role to describe our Universe.  The Chaplygin gas 
\cite{Kamenshchik:2001cp,Bilic:2001cg} and its generalization \cite{Bento:2002ps} will be of special interest for us in this work.

In this paper we find first that the different epochs of the standard cosmological model can be described in the context of GTD and that these correspond to 
the simplest case of a system with no internal thermodynamic interaction. Thereafter, we consider a second GTD system with thermodynamic interaction that turns out to 
describe a unified dark sector fluid which has as a special case the generalized Chaplygin gas. As a bonus, the so-called polytropic fluids can be obtained from 
this GTD fluid in a certain limit.

This paper is organized as follows. In Sec. \ref{sec:gtd}, we briefly review the fundamentals of GTD. Then, in Sec. \ref{sec:scm}, we present the cosmological model that follows from a GTD system without thermodynamic interaction. 
In Sec. \ref{sec:dark}, we study the cosmology of the dark sector GTD fluid at the homogeneous and isotropic level. In Sec. \ref{sec:darkpert},
we work out the linear perturbation theory in order to constrain the free parameters of the GTD model with thermodynamic interaction. Finally, section \ref{sec:con} 
is devoted to discussions of our results and suggestions for further research. Throughout this paper we use geometric units in which $G=c=k_{_B}=\hbar =1$, unless otherwise stated.

\section{Basic aspects of geometrothermodynamics}
\label{sec:gtd}

In classical equilibrium thermodynamics \cite{Callen:Thermodynamics}, the standard method to describe a thermodynamic system consists in 
specifying a set of $n$ extensive variables $E^a$ ($a=1,...,n$), their corresponding dual intensive variables $I^a$, and 
the thermodynamic potential $\Phi$. The integer $n$ determines the number of thermodynamic degrees of freedom of the system. 
For instance, in the case of the ideal gas $(n=2)$, if we  choose the internal energy $U$ as the thermodynamic potential $\Phi$, 
then $E^a=(S,V)$ and $I^a=(T,-P)$ so that the temperature $T$ is  the dual of the entropy $S$ and  the (negative) pressure $P$ 
is the dual of the volume $V$. All the properties of the ideal gas are contained in the fundamental equation $U=U(S,V)=(e^S/V)^{2/3}$ 
that satisfies the first law of thermodynamics $dU = TdS - P dV$ from which the expressions for the temperature and the pressure, i.e., 
the equations of state can be derived. Since an equilibrium state of the ideal gas can be represented by the corresponding values of $S$ 
and $V$, all possible equilibrium states form a space ${\cal E}$ whose points can be represented by the coordinates $S$ and $V$. 

Notice that using the above notation for an arbitrary system with $n$ thermodynamic degrees of freedom, the fundamental equation can 
be written as $\Phi=\Phi(E^a)$,  the first law of thermodynamics as $d\Phi = I_a d E^a$ with $I_a=\delta_{ab}I^b$, and the coordinates 
of the equilibrium space ${\cal E}$ are $E^a$. An advantage of this notation is that it can be used with any thermodynamic potential 
and representation. For instance, to write the above example of the ideal gas in the entropy representation one only needs to rewrite 
the first law of thermodynamics as $dS=(1/T)dU + (P/V) dV$ so that the thermodynamic variables are now $\Phi=S$, $E^a=(U,V)$, 
and $I^a=\left(1/T, P/T \right)$.  

An important property of classical thermodynamics is that it is invariant with respect to Legendre transformations, i.e., it does not 
depend on the choice of thermodynamic potential. Indeed, for the description of the ideal gas instead of $U$ one can also use as 
thermodynamic potential the Helmholtz free energy $F=U-TS$, the enthalpy $H=U+PV$ or the Gibbs energy $G=U-TS+PV$, without changing 
the properties of the system. The Legendre transformations that generate the potentials $F$ and $H$ are called partial transformations 
whereas $G$ is generated by a total transformation. 

The main idea of GTD consists in associating a differential geometric structure to the equilibrium space of a given thermodynamic system 
in such a way that it does not depend on the choice of the thermodynamic potential, i.e., it is Legendre invariant. To this end, it is 
necessary to introduce an auxiliary structure called the phase space in which the equilibrium space is embedded. 
To be more specific, let us define the phase space as the  
$(2n+1)-$dimensional differential manifold
${\cal T}$, with coordinates $Z^A=\{\Phi, E^a, I^a\}$, $A=0,...,2n$, equipped with the fundamental Gibbs one-form 
$\Theta = d\Phi - I_a dE^a$ \cite{Hermann},  and a metric $G$ that must be invariant with respect to Legendre transformations. 
The last condition is necessary in order to incorporate in GTD the fact that classical thermodynamics is Legendre invariant. 
In this notation, a Legendre transformation is given by 
\begin{equation}
 \{ Z^A \} \rightarrow \{ \tilde{Z}^A \} = \{ \tilde{\Phi}, \tilde{E}^a, \tilde{I}^a\},
\end{equation}
with
\begin{equation}
\Phi = \tilde{\Phi} - \delta_{kl}\tilde{E}^k\tilde{I}^l, \quad E^i = - \tilde{I}^i, \quad I^i = \tilde{E}^i, \quad E^j = \tilde{E}^j, \quad I^j = \tilde{I}^j.
\end{equation}
Here $i,k,l \in I$ and $j \in J$, where $I \cup J$ is any disjoint decomposition of the set of indices $\{1, \dots, n \}$.
The metric \cite{Vazquez:2011ia} (summation over all repeated indices)
\be
G= (d\Phi - I_a dE^a)^2 + \Lambda \, E_a I_a dE^a dI^a \ 
\ee
where  $\Lambda$ is a real constant, is the most general metric we have found so far that is invariant under partial and total Legendre transformations, 
and the last term linear in the extensive and intensive variables. 

The equilibrium submanifold ${\cal E} \subset {\cal T}$
is defined by the smooth map $\varphi: {\cal E}\rightarrow {\cal T}$, or in coordinates $\varphi: \{ E^a\}\mapsto  \{\Phi(E^a), E^a, I^a(E^a)\}$, 
under the condition that $\varphi^*(\Theta)=0$, i.e.,
\be
\label{flaw}
d\Phi =  I_a dE^a\ ,\quad {\rm i.e.,}\quad I_a = \frac{\partial \Phi}{\partial E^a}\ ,
\ee
where $\varphi^*$ is the pullback of $\varphi$. These equations are equivalent to the first law of thermodynamics and the conditions for thermodynamic equilibrium, respectively. 
We can associate with ${\cal E}$ the induced metric 
\be
\label{gdown}
g=\varphi^*(G) = \Lambda \left(E_a \frac{\partial \Phi}{\partial E^a}\right) \frac{\partial^2\Phi}{\partial E^b \partial E^c}\delta^{ab} dE^a dE^c
\ee
in a canonical manner. 
One of the main objectives of GTD is to find 
relations between the geometric properties of the equilibrium space ${\cal E}$ and the thermodynamic properties of the system determined by the 
fundamental equation $\Phi=\Phi(E^a)$ \cite{Callen:Thermodynamics} that, in turn, is specified by the map $\varphi$. In particular, one expects that the curvature 
of ${\cal E}$ can be used as a measure of the thermodynamic interaction. For instance, in the case of vanishing interaction, one 
expects the curvature to be zero.
Let us recall that our interpretation of thermodynamic interaction is based upon the statistical approach 
to thermodynamics in which all  the properties of the system can be derived from the explicit form of the corresponding Hamiltonian \cite{Greiner:Thermodynamics},
and the interaction between the particles of the system is described by the potential part of the Hamiltonian. Consequently, if the potential vanishes, 
we say that the system has zero thermodynamic interaction and the curvature should vanish. The equivalence between the curvature of ${\cal E}$ and the thermodynamic interaction 
has been shown to be true in the case of ordinary classical systems, like 
the ideal gas and the van der Waals gas \cite{Vazquez:2011ia}, and black hole configurations in different theories (see \cite{Quevedo:2010nv} for a review). 
Moreover, the curvature singularities of ${\cal E}$ turn out to correspond to phase transitions of the thermodynamic system.

The above description of GTD shows that in order to find explicitly the metric $g$ of the equilibrium manifold ${\cal E}$ one only needs to specify the 
fundamental equation $\Phi= \Phi(E^a)$. This means that one needs the fundamental equation to study the corresponding geometry. However, the formalism of 
GTD allows us to generate fundamental equations by using a variational principle as follows. Suppose  that the equilibrium manifold ${\cal E}$ determines 
an extremal surface in ${\cal T}$, i.e., the variation of the volume element of ${\cal E}$ vanishes: 
\begin{equation}
 \delta \int_{\cal E} \sqrt{{\rm det}(g)} d^n E =0.
\end{equation}
Since $g$ is induced by the metric $G$ that depends on $Z^A$, it can be shown \cite{Vazquez:2011ia} that this variation leads to a system of differential 
equations 
\begin{equation}
 \Box Z^A = \frac{1}{\sqrt{{\rm det}(g)}}\left( \sqrt{{\rm det}(g)} g^{ab} Z^A_{,a} \right)_{,b} + \Gamma^{A}_{\, BC} Z^B_{,b} Z^C_{,c} g^{bc}=0
\end{equation}
where $\Box$ is the d'Alembert operator. Moreover, this variation implies that the thermodynamic potential $\Phi$ 
must satisfy a set of differential equations whose solutions can be written as functions of the extensive variables $\Phi=\Phi(E^a)$, i.e., as fundamental 
equations.  Two particularly simple 
solutions with $\Phi=S$ and $E^a=\{U,V\}$ found in \cite{Vazquez:2011ia} are given by

\begin{equation}
S= c_1 \ln U + c_2 \ln V, \label{varproEq1}
\end{equation}
and
\begin{equation}
S=S_0 \ln \left( U^{1+\alpha} + c V^{1+\beta}\right), \label{varproEq2}
\end{equation}
where $c_1$, $c_2$, $\alpha$ and $\beta$ are real constants.

The question arises whether these functions, which are obtained as solutions of a geometric problem, can be used as fundamental equations to 
describe a thermodynamic system with realistic physical properties. This question will be treated in the following sections.

\section{The fluids of the standard cosmological model}
\label{sec:scm}

The simplest solution with two thermodynamic degrees of freedom ($n=2$) is given by equation (\ref{varproEq1}) In the special case $c_1=3/2$ and $c_2=1$, 
we obtain the Sackur-Tetrode equation that is interpreted as the fundamental equation for the ideal gas \cite{Callen:Thermodynamics}. This solution is the simplest one in 
the sense that it corresponds to a system with no internal thermodynamic interaction. In fact, introducing the eq. (\ref{varproEq1}) into the general 
metric (\ref{gdown}) with $\Phi=S$ and $E^a=\{U,V\}$, 
 we obtain the particular metric
\be
g= - \Lambda\left( c_1^{2} \frac{dU^2}{U^2} + c_2^{2} \frac{dV^2}{V^2}\right)\ .
\ee
A straightforward calculation shows that the curvature of this metric vanishes identically, showing that the metric is flat. This can be seen 
explicitly by introducing the coordinates $d\xi=\Lambda^{1/2} c_1 dU/U$ and $d\eta = \Lambda^{1/2} c_2 dV/V$ in which the metric takes the 
Euclidean form $g= -(d\xi^2 + d\eta^2)$. As mentioned above, in GTD we interpret the curvature as a measure of the thermodynamic interaction so 
that a flat metric corresponds to the simplest case of a system without interaction.

The first law of thermodynamics (\ref{flaw}) in the entropy representation can be written as 
\be
dS=\frac{1}{T} dU + \frac{P}{T} dV \ . 
\ee
Then, from the equilibrium conditions (\ref{flaw}) we obtain the relationships $T=U/c_1$ and $P/T= c_2/V$ which lead to the equation of state
\be
\label{eosbar} 
P=\frac{c_2}{c_1} \rho \ ,
\ee 
where $\rho = U/V$. To consider this thermodynamic system in general relativity we assume the simplest case of a homogeneous and isotropic 
spacetime that is described by the Friedmann-Lem\^aitre-Robertson-Walker (FLRW) line element
\be
\label{lineelement}
ds^2 = -dt^2 + a(t)^2\left[\frac{dr^2}{1-kr^2} + r^2 \left(d\theta^2 +\sin^2\theta d\phi^2\right)\right]\ .
\ee
Then, if we assume a perfect fluid source with equation of state (\ref{eosbar}), it is clear that the different epochs of the Universe evolution 
can be obtained by choosing the constants appropriately. So, the choice $c_2/c_1=1/3$ corresponds to the radiation dominated era, $c_2=0$ describes 
the matter dominated era, and $c_2/c_1=-1$ corresponds to a vacuum dominated cosmology.  Consequently, the different fluids of the standard model 
can be described by applying the simplest GTD fundamental equation (\ref{varproEq1}) in the context of general relativity; in other words, the fluids of 
the standard cosmological model correspond thermodynamically to the simplest possible fundamental equation of GTD. 

It is worth noticing that for the fundamental equation (\ref{varproEq1}) the heat capacity at constant volume is given by $C_V=c_1$. This opens the 
possibility of considering the dark energy as a non-interacting thermodynamic system with negative heat capacity. In fact, for the dark energy 
fluid we obtained that $c_2/c_1=-1$; therefore, we can assume that $c_2>0$ which results in a negative $C_V$. Although most physical systems 
exhibit a positive heat capacity, there are systems for which the heat capacity is negative. Among others, these include self-gravitating objects 
such as stars and star clusters \cite{springerlink:10.1007/BF01403177}. Furthermore, it can be shown \cite{LyndenBell:1998fr}
that systems with negative $C_V$ are never extensive. We conclude that the dark energy fluid can be considered as a non-interacting system with 
non-extensive thermodynamic variables. To further investigate this possibility it is necessary to consider non-extensive variables in the framework 
of GTD. We expect to study this problem in the near future.

\section{A unified description for dark matter and dark energy} 
\label{sec:dark}

In this section we study the fundamental equation (\ref{varproEq2}). According to eq.(\ref{gdown}), this solution generates the thermodynamic metric
\bea
g &= &\frac{\Lambda S_0^2}{(U^{1+\alpha}+cV^{1+\beta} )^3} \bigg[
(1+\alpha)^2U^{2\alpha} [{\alpha c V^{1+\beta} - U^{1+\alpha}}]dU^2  \nonumber \\
&+& (1+\beta)^2c^2 V^{2\beta}[ {\beta U^{1+\alpha}-cV^{1+\beta}} ] dV^2 \nonumber \\
&-& (1+\alpha)(1+\beta)cU^{1+\alpha} V^{1+\beta}[(1+\alpha)U^{1+\alpha}+(1+\beta)cV^{1+\beta}]dUdV \bigg]\ , 
\label{gdown2}
\eea
for the equilibrium manifold ${\cal E}$. The corresponding curvature is, in general, non-vanishing and in the particular case $\alpha=1$ and $\beta=1$ it can be expressed as
\be
R= 
{\frac {6\,{U}^{4}{V}^{4}{c}^{2}+4\,{U}^{6}{V}^{2}c+4\,{U}^{2}{V}^{6}{
c}^{3}+{V}^{8}{c}^{4}+{U}^{8}}{{S_0}^{2} \left( {c}^{2}{V}^{4}+{U
}^{4} \right) ^{2}}}
\ ,
\ee
indicating the presence of thermodynamic interaction.
In this sense, this thermodynamic system represents a generalization of the system with no interaction investigated in the last section. 
The first law of thermodynamics is again (\ref{flaw}) and the conditions of thermodynamic equilibrium lead to
\be
\frac{1}{T}= \frac{S_0(1+\alpha) U^\alpha}{ U^{1+\alpha} + c V^{1+\beta} } \ ,\quad
\frac{P}{T} = \frac{S_0 c (1+\beta ) V^\beta}{ U^{1+\alpha} + c V^{1+\beta} }\ .
\ee
Then, an equation of state can be written as
\begin{equation}
P(U,V) =  \frac{c(1+\beta) V^{\beta}}{(1+\alpha)U^{\alpha}} \label{DSEoS} \ .
\end{equation}

We now consider the large scale evolution of a universe filled with standard model particles and the dark sector described by GTD; 
a subindex $d$ shall denote  GTD dark, single fluid, variables. We write the equation of state of the dark sector (\ref{DSEoS}) in terms of the scale factor 
$a(t)$ and its energy density $\rho_d$
\begin{equation}
 P_d = -\mathcal{C} a^{-3(\alpha-\beta)} \rho_d^{-\alpha} \label{DSEoS2}
\end{equation}
where we used $V = V_0 (a/a_0)^3$. Also, we set the value of the scale factor today equal to one, and defined the constant $\mathcal{C}= -c (1+\beta) V_0^{\beta - \alpha}/ (1+\alpha)$.
From this equation one can see that the specific case $\alpha=\beta$ in the interval $0 \le \alpha \le 1$ corresponds to a (generalized) Chaplygin gas 
\cite{Kamenshchik:2001cp,Bento:2002ps}. Moreover, if 
$\alpha=\beta=0$ a fluid, often called  {\it dark fluid}, which gives exactly the same phenomenology as the $\Lambda$CDM model is obtained 
\cite{Avelino:2003cf,Aviles:2011ak,Luongo:2011yk}, not only at the cosmological level, but also at astrophysical scales. This is because the dark fluid
which comprises about $96 \%$ of the energy content of the Universe, partially clusters; for details see \cite{Kunz:2007rk,Aviles:2011ak}.

Polytropic fluids, extensively used in modeling astrophysical objects, are obtained if $\alpha=\beta$ in the interval $\alpha < 0$.
We also note that the case $\alpha = 1$, dubbed {\it variable Chaplygin gas}, has been studied in the past and has the advantage over the standard Chaplygin that it 
can develop large inhomogeneous perturbations \cite{Guo:2005qy,Bilic:2008yr}.
 
It is interesting to note that the dark fluid model with $\alpha=\beta=0$
leads to a thermodynamic metric (\ref{gdown2}) whose curvature vanishes identically. This resembles the case of the GTD fluid described in section \ref{sec:scm} 
that generates the fluids of the standard cosmological model.

The continuity equation $\rho_d' = - 3 H (\rho_d + P_d)$ (prime denotes differentiation with respect to cosmic time, contrary to conformal time, to be used in the next section)
can be integrated to give

\begin{equation}
\rho_d = \left[\frac{1+\alpha}{1+\beta}\mathcal{C} a^{-3(\alpha-\beta)} + C_I a^{-3(1+\alpha)} \right]^{1/(1+\alpha)},
\end{equation}
where $C_I$ is an integration constant. It is convenient to recast this expression into the form
\begin{equation}
\rho = \rho_{d0}  \left( \mathcal{A} a^{-3(\alpha-\beta)} + (1-\mathcal{A}) a^{-3(1+\alpha)} \right)^{1/(1+\alpha)}\ ,
\end{equation}
where we defined $\rho_{d0}$ as the  value of the dark sector energy density today. The constants are related by the equations
\begin{eqnarray}
 \mathcal{A} = \frac{\mathcal{C}}{\mathcal{C} + C_I (1+\beta) /(1+\alpha)}, & \qquad & \rho_{d0}  = \left( \frac{1+\alpha}{1+\beta} \mathcal{C} + C_I \right)^{1/(1+\alpha)} \nonumber \\
\mathcal{C} = \frac{1+\beta}{1+\alpha}\mathcal{A}\,\rho_{d0}^{1+\alpha},  & \qquad & C_I = \rho_{d0}^{1+\alpha} \left( 1 -  \mathcal{A} \right). 
\end{eqnarray}
To ensure the reality and positivity of $\rho_d$ at all times, we must impose the condition $\mathcal{A}>0$ that implies the relation $c(1+\beta)/(1+\alpha)<0$. 
Notice that for $1+\alpha<0$ and positive $S_0$, it follows that  $\partial S/ \partial U <0$ and so the possibility of a negative heat capacity arises, as in 
the case analyzed in the previous section. We will not investigate this case in this section. Thus, following eq. (\ref{varproEq2}), the entropy of 
the system must diminish as the configuration space grows,  and as a consequence the GTD dark fluid has a negative pressure 
which ultimately is responsible to accelerate the Universe.
    
\begin{figure}[ht]
	\begin{center}
	\includegraphics[width=3.3 in]{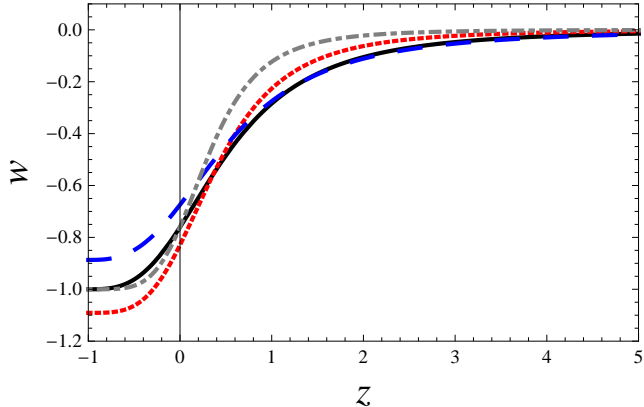}
	{\small \caption{Evolution of $w_d$ as a function of the redshift $z$. The solid (black) curve corresponds to $\alpha=\beta=0$ (the $\Lambda$CDM model); 
          the dashed (blue) curve to $\alpha=0.06$ and $\beta=-0.06$; the dotted (red) curve to  $\alpha=0.1$ and $\beta=0.2$; the dash-dotted (gray) curve to  
          $\alpha=\beta=0.5$ (a Chaplygin gas). $\mathcal{A}= 0.76$ is kept fixed for all the cases.}
	\label{fig:EoS}}
	\end{center}
\end{figure}

It is straightforward to calculate the equation of state parameter of the GTD dark sector fluid ($w_d=P_d/\rho_d$), giving

\begin{equation}
 w_d(a) = -\frac{1+\beta}{1+\alpha} \,\frac{1}{1 + (1-\mathcal{A}) a^{-3(1+\beta)}/\mathcal{A}}, \label{wda}
\end{equation}
which has the following behavior
\begin{eqnarray}
 w_d(a\rightarrow 0) &\quad \longrightarrow \quad& \quad \quad \! 0, \nonumber\\
w_d(a \rightarrow \infty ) &\quad \longrightarrow \quad& - \frac{1+\beta}{1+\alpha} , \\
w_d(a=1)  & = & - \frac{1+\beta}{1+\alpha} \, \mathcal{A}.
\end{eqnarray}
Figure \ref{fig:EoS} shows the evolution of $w_d$ as a function of the redshift $z = 1/a -1$ for different combinations of $\alpha$ and $\beta$; $\mathcal{A}$ is kept fixed to the value 
$\mathcal{A} = 1/(1+ \Omega_{DM}/\Omega_{\Lambda}) \simeq 0.76$,  with $\Omega_i = 8 \pi \rho_{i0}/ 3 H_0^2$.

Now, the Friedmann equation is given by
\begin{equation}
 H^2 = \frac{8 \pi}{3}(\rho_d + \rho_b + \rho_{\gamma}), \label{s1}
\end{equation}
where $H \equiv a'/a$ is the Hubble factor. The energy densities of baryons ($\rho_b$) and relativistic components ($\rho_\gamma$) redshift as
$\rho_b = \rho_{b0} a^{-3}$ and $\rho_{\gamma} = \rho_{\gamma 0} a^{-4}$, respectively. 

\begin{figure}[ht]
	\begin{center}
	\includegraphics[width=3.3 in]{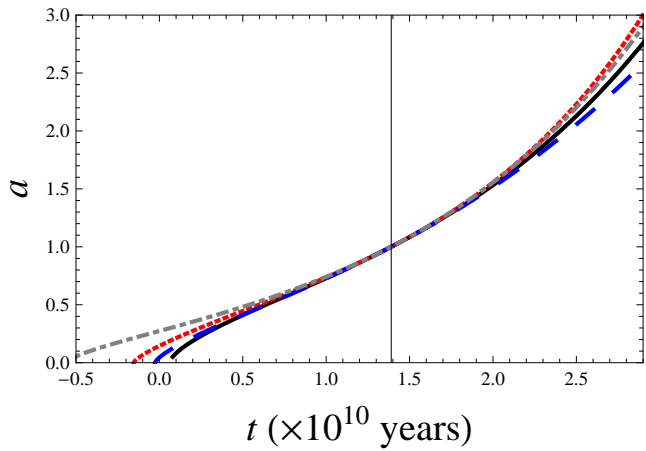}
	{\small \caption{Evolution of the scale factor $a$ as a function of the cosmic time $t$. The solid (black) curve corresponds to $\alpha=\beta=0$ 
          (the $\Lambda$CDM model); the dashed (blue) curve to $\alpha=0.06$ and $\beta=-0.06$; the dotted (red) curve to  $\alpha=0.1$ and $\beta=0.2$; 
           the dash-dotted (gray) curve to  $\alpha=\beta=0.5$. $\mathcal{A}= 0.76$ 
           and $\Omega_d = 0.96$ are kept fixed for all the cases. The vertical line denotes present time.}
	\label{fig:avst}}
	\end{center}
\end{figure}

To complete with the homogeneous and isotropic description we solve numerically the Friedmann equation. We choose the same values for $\alpha$, $\beta$ and $\mathcal{A}$ 
as in figure 1.
The value of $\rho_{d0}$ is fixed by the flat condition, $\Omega_d + \Omega_b + \Omega_r = 1$, giving $\Omega_d \simeq 0.96$.
In figure \ref{fig:avst} we plot the scale factor as a function of the cosmic time for the different chosen combinations of the parameter values.

An important quantity for the investigation of the fluid perturbations ---to be analyzed in the next section--- but calculated with purely background quantities 
is the square of the adiabatic speed of sound, $c^2_s \equiv \dot{P}_d/\dot{\rho}_d$, which can be shown to be

\begin{equation}
 c^2_s = - w_d \frac{\alpha P_d + \beta \rho_d}{\rho_d + P_d},
\end{equation}
or, written as a function of the scale factor,
\begin{equation}
 c^2_s = \frac{1+\beta}{1+\alpha} \,\frac{1}{1 + (1-\mathcal{A}) a^{-3(1+\beta)}/\mathcal{A}} 
\frac{(\beta-\alpha)/(1+\alpha) + \beta (1-\mathcal{A}) a^{-3(1+\beta)}/\mathcal{A} }{(\alpha-\beta)/(1+\alpha) +  (1-\mathcal{A}) a^{-3(1+\beta)}/\mathcal{A} }. \label{cs2}
\end{equation}
The limits of this expression are $c_s^2(a\rightarrow 0) = 0$, and $c_s^2(a\rightarrow \infty ) = - (1+\beta)/(1+\alpha)$, if $\alpha \neq \beta$, and
$c_s^2(a\rightarrow \infty ) = \alpha$, if $\alpha = \beta$. This  result leads to an important difference between the generalized Chaplygin model and the extension
found here with GTD. At the cosmological background level this fact does not have any consequences, but as we shall see, 
it is of great importance when considering perturbations. To not violate causality we require $c^2_s \le 1$; consequently, further conditions are imposed over 
the parameters $\alpha$ and $\beta$.

The particular case of the Chaplygin gas gives $c^2_s = -\alpha w_d$, while for the dark fluid, $c^2_s = 0$. 
The assumption that the speed of sound vanishes has been the starting point in several works that study the dark fluid model as an alternative 
to the $\Lambda$CDM \cite{Luongo:2011yk,Aviles:2011ak}. It turns out that both models are fundamentally 
indistinguishable  as long as some general conditions are imposed beyond the zero order in perturbation theory. Instead of the Chaplygin gas, it is possible to consider 
its natural extension based upon a constant speed of sound, an approach adopted in  \cite{Balbi:2007mz,Xu:2011bp}.

We note that if $\alpha < \beta$, there is a singularity in the speed of sound at $a = [(1+\alpha) (1-\mathcal{A})/\mathcal{A} (\beta - \alpha)]^{1/3(1+\beta)}$, 
this coincides with the moment at which the equation of state parameter crosses the phantom barrier, $w_d = -1$.

\section{The perturbed Universe}
\label{sec:darkpert}

At small scales (nowadays lesser than about $100 \,\text{Mpc}$) the homogeneous and isotropic description of the Universe outlined in the last section 
breaks down. In this section we study the deviations of the background cosmology up to linear order in perturbation theory. To this end, let us consider 
scalar perturbations in the Conformal Newtonian gauge, with the line element given by 

\begin{equation}
 ds^2 = a^2(\tau)\big[ -(1+2 \Psi) d\tau^2 + (1- 2 \Phi) \delta_{ij} dx^i dx^j \,\big],
\end{equation}
where $\tau$ is the conformal time, related to the cosmic time by $dt = a d \tau$.
The matter fields perturbation variables are defined through the expressions

\begin{eqnarray}
T^{0}{}_{0} &=& -\rho(1+ \delta), \\
T^{i}{}_{0} &=& - (\rho + P) v^i, \\
T^{i}{}_{j} &=& P \big( (1+\pi_L) \delta^{i}{}_{j} + \Pi^{i}{}_{j} \big),
\end{eqnarray}
where  $\Pi^{i}{}_{j}$ is the  anisotropic stress tensor. The energy density $\rho$ and the pressure $P$ denote background quantities, and are 
functions of the conformal time only. The vector $v^i$ is called the peculiar velocity and is related to the four-velocity $u^{\mu}$ of the 
fluid by the relation $v^i = u^i / u^0$. In the Fourier space we define the velocity $\theta = - i k_i v^i$ and the scalar anisotropic  stress
$\sigma = 2 k_i k_j \Pi^{ij} w / 3 (1+w)$. 

For a general fluid the energy local conservation equations $\nabla_{\mu}T^{\mu\nu}= 0$ become \cite{Ma:1995ey}

\begin{equation}
 \dot{\delta} = - (1+ w)(\theta - 3 \dot{\Phi}) - 3 \mathcal{H} \left( \frac{\delta P}{\delta \rho} - w \right) \delta,  \label{pertd} \\
\end{equation}
and
\begin{equation}
 \dot{\theta} = - \mathcal{H} (1 - 3 w)\theta - \frac{\dot{w}}{ 1+w} \theta + \frac{ \delta P / \delta \rho}{1+w}k^2 \delta 
              + k^2 \Psi  -  k^2 \sigma,  \label{pertt} 
\end{equation}
where  $\delta P = P \pi_L$, $\delta \rho = \rho \delta$, $\mathcal{H} = \dot{a}/a$ and a dot means derivative with respect to conformal time. Note
that the adiabatic speed of sound can be expressed as $c^2_s = w - \dot{w} /3\mathcal(1+w)$.
To go further on, we make the assumption of a perfect fluid, obtaining no anisotropic stresses, $\sigma = 0$, so that the gravitational potentials 
coincide, $\Phi = \Psi$. Moreover, if we consider only adiabatic perturbations, then the (gauge invariant) entropy perturbation is zero, 
$\Gamma = \pi_L - c^2_s \delta / w = 0$, and  the equations for the GTD dark sector fluid become

\begin{equation}
 \dot{\delta}_d = - (1+ w_d)(\theta_d - 3 \dot{\Phi}) - 3 \mathcal{H} \left( c^2_s - w_d  \right)\delta_d ,  \label{pertd2} \\
\end{equation}
and
\begin{equation}
 \dot{\theta}_d = - \mathcal{H} (1 - 3 c^2_s)\theta_d  + \frac{ c^2_s k^2 \delta_d}{1+w_d} + k^2 \Phi,  \label{pertt2} 
\end{equation}
where $w_d$ and $c^2_s$ are given by eqs. (\ref{wda}) and (\ref{cs2}), respectively. 

\begin{figure}[ht]
	\begin{center}
	\includegraphics[width=3 in]{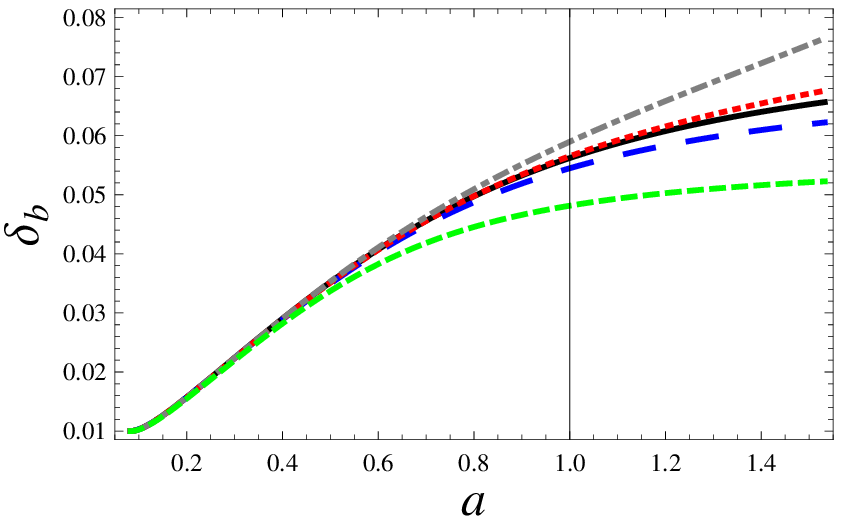}
        \includegraphics[width=3 in]{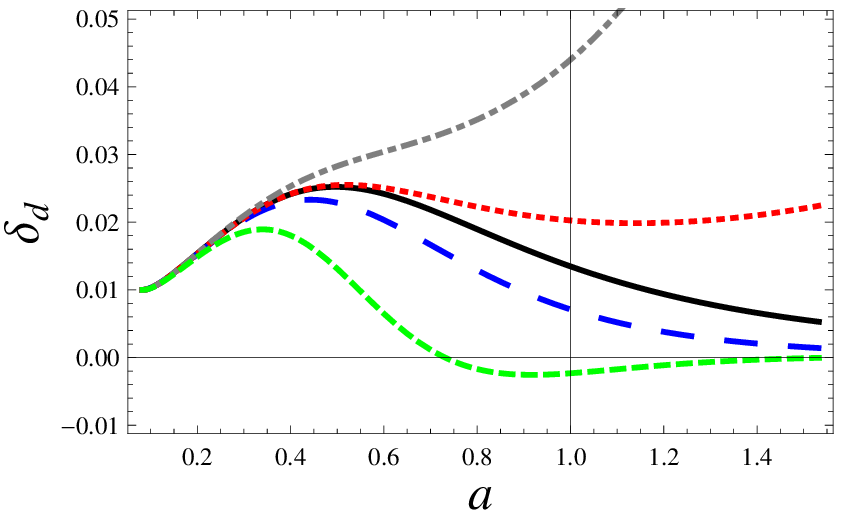}
	{\small \caption{Evolution of the baryonic (left panel) and dark sector fluid (right panel) density contrasts as a function of the scale factor $a$. 
                        Solid (black) line corresponds to $\alpha=\beta=0$ (the dark fluid model). Large-dashed (blue) line to 
                        $\alpha=\beta = 0.0001$. Short-dashed (green) line to  $\alpha=\beta= 0.0006$. Dash-Dotted (gray) line to  $\alpha=0.0001$ and $\beta=-0.0001$.
                        $\mathcal{A}= 0.76$ and $\Omega_d = 0.96$ are kept fixed for all the cases.}
	\label{fig:deltas}}
	\end{center}
\end{figure} 

\begin{figure}[ht]
	\begin{center}
	\includegraphics[width=3.5 in]{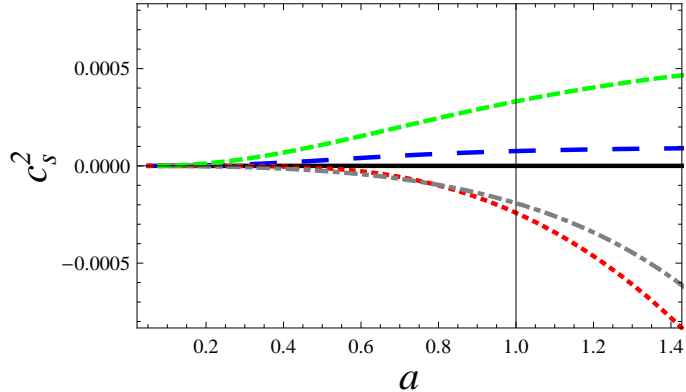}
	{\small \caption{Evolution of the adiabatic speed of sound for the cases considered in figure 4.}
	\label{fig:cs22}}
	\end{center}
\end{figure} 

Figure \ref{fig:deltas} shows the behavior of the baryons and dark sector density contrasts, $\delta_b$ and $\delta_d$ respectively, for different chosen parameters 
$\alpha$ and $\beta$. We note that in the cases with $\alpha = \beta$ (Chaplygin gases), the density contrasts decay more 
quickly than those with $\alpha \neq \beta$ (not Chaplygin gases). This is because, as shown in figure \ref{fig:cs22}, the squared of the speed of sound of the perturbations 
is positive for the former cases  and negative for the latter, enhancing the growth of structure. See eq. (\ref{cs2}) and the discussion thereafter.

\begin{figure}[ht]
	\begin{center}
	\includegraphics[width=3.5 in]{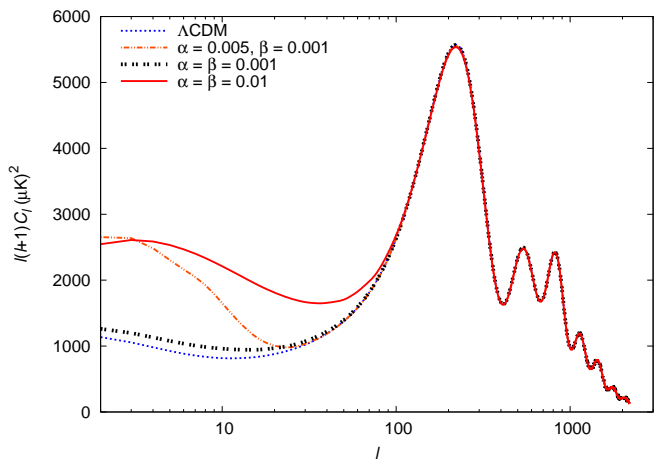}
	{\small \caption{CMB angular power spectrum for different values of $\alpha$ and $\beta$. $\mathcal{A} = 0.76$ is kept fixed.}
	\label{fig:CMB}}
	\end{center}
\end{figure}

To proceed with the analysis we use the publicly available code CAMB \cite{Lewis:1999bs} to study the anisotropies of the cosmic background radiation. 
In figure \ref{fig:CMB} we plot the CMB angular power spectrum for different choices of the parameters $\alpha$ and $\beta$, keeping fixed the remaining parameters.
We note that the larger deviations from the $\Lambda$CDM model show up at large scales. This can easily be understood from the equation of state parameter 
and the adiabatic speed of sound: both of them are nearly zero at high redshifts, thus at early times the GTD dark sector fluid behaves essentially as 
cold dark matter, then at lower redshifts ---after recombination for the cases shown in figure \ref{fig:CMB}--- they start to diverge from the zero values. 
Consequently, the differences arise mainly through the integrated Sachs-Wolfe effect. This enhancement of the low CMB power spectrum multipoles has been 
found in the past for the Chaplygin gas \cite{Amendola:2003bz}, and in general for unified dark models  \cite{Bertacca:2007cv}.

To constrain the parameters of the model, we use the code CosmoMC \cite{Lewis:2002ah} to perform a Markov Chain Monte Carlo (MCMC) analysis 
over the eight-parameter space $\mathcal{M} = \{\Omega_b h^2, \theta, \tau, n_s, \log A_s, \alpha, \beta, \mathcal{A} \}$.
$\theta$ is defined as 100 times the ratio of the sound horizon to the angular diameter distance at recombination,
$\tau$ is the reionization optical depth,  $n_s$ is the spectral index of the primordial scalar perturbations and $A_s$ is its amplitude at a pivot scale of 
$k_0 = 0.05\, \text{Mpc}^{-1}$. We take flat priors on the intervals $-0.01< \alpha, \beta < 0.02$ and $0.2 < \mathcal{A} < 0.99$.

The observations that we choose to constrain the model are the WMAP seven-years results of the observations of the anisotropies of the CMB 
\cite{Larson:2010gs}, and the supernovae type Ia Union 2 data set compilation of the Supernovae Cosmology Project \cite{Amanullah:2010vv}. 
Moreover, we use Hubble Space Telescope (HST) measurements to impose a Gaussian prior on the present value of the Hubble constant of 
$H_0 = 74 \pm 3.6 \,\text{km/s/Mpc}$ \cite{Riess:2009pu}.

\begin{figure}[ht]
	\begin{center}
	\includegraphics[width=2.5 in]{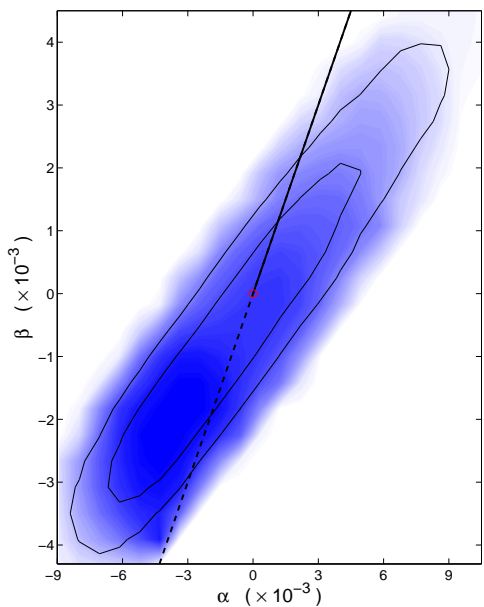}
	{\small \caption{Contour confidence intervals for the $\alpha - \beta$ subspace of parameters at $68 \%$ and $95 \%$ c.l. The solid line
                         corresponds to the generalized Chaplygin gas, the dashed to a polytropic fluid and the circle is the $\Lambda$CDM model.
                         The shading shows the mean likelihood of the samples.}
	\label{fig:alphavsbeta}}
	\end{center}
\end{figure}

Figure \ref{fig:alphavsbeta} shows the marginalized confidence interval in the subspace $\alpha-\beta$; in this figure, the region of parameters that
corresponds to the Chaplygin gas is represented by a solid straight line, and the polytropic case by a dashed line. 
These lines split the space into two regions, 
$\alpha> \beta$ (with no singular solutions) and $\alpha < \beta$. The circle corresponds to the dark fluid (or $\Lambda$CDM) model. 

\begin{figure}
\begin{center}
\includegraphics[width=1.9in]{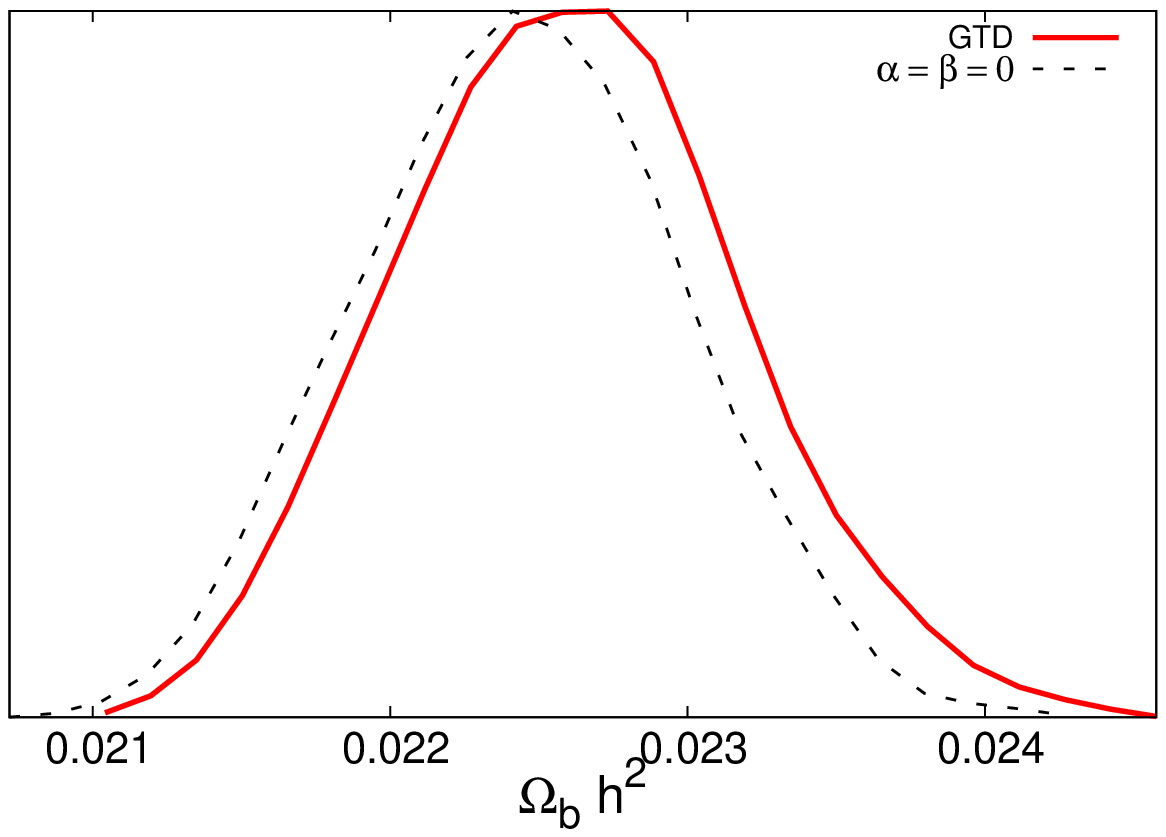}
\includegraphics[width=1.9in]{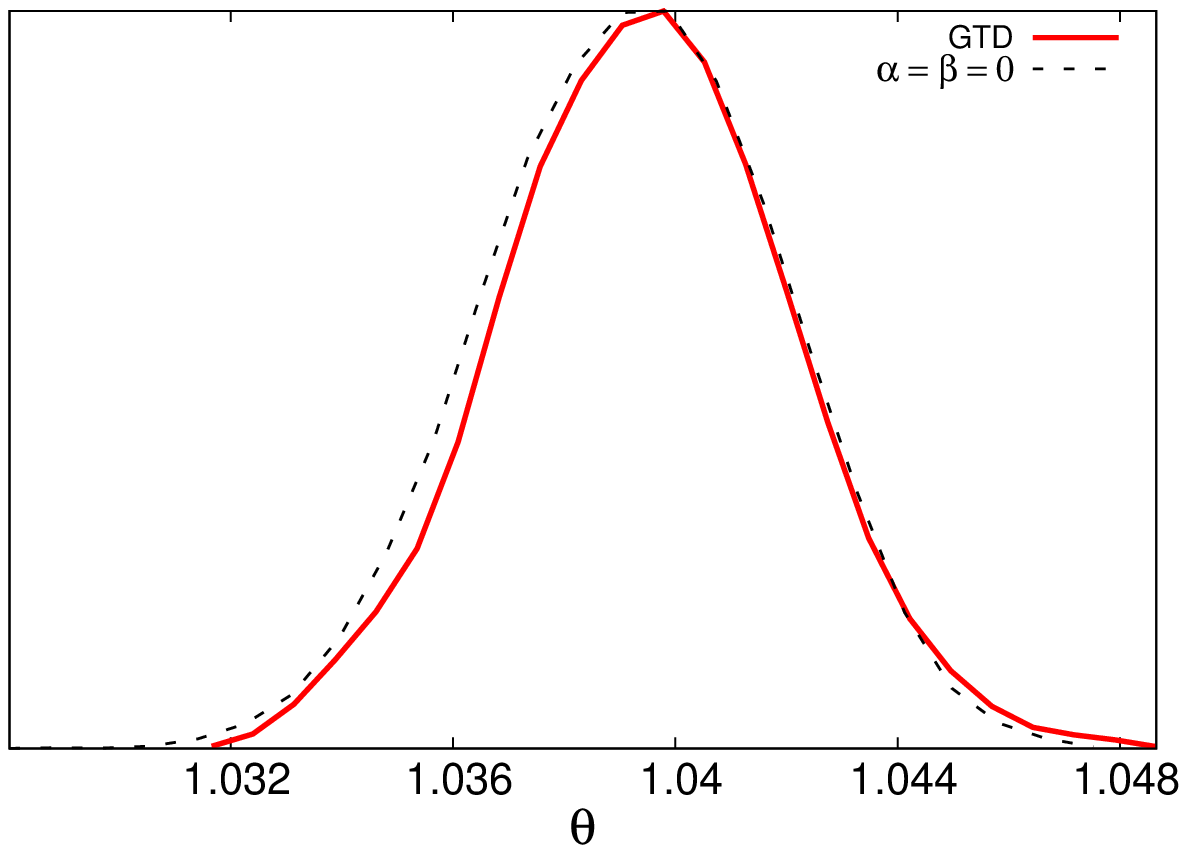}
\includegraphics[width=1.9in]{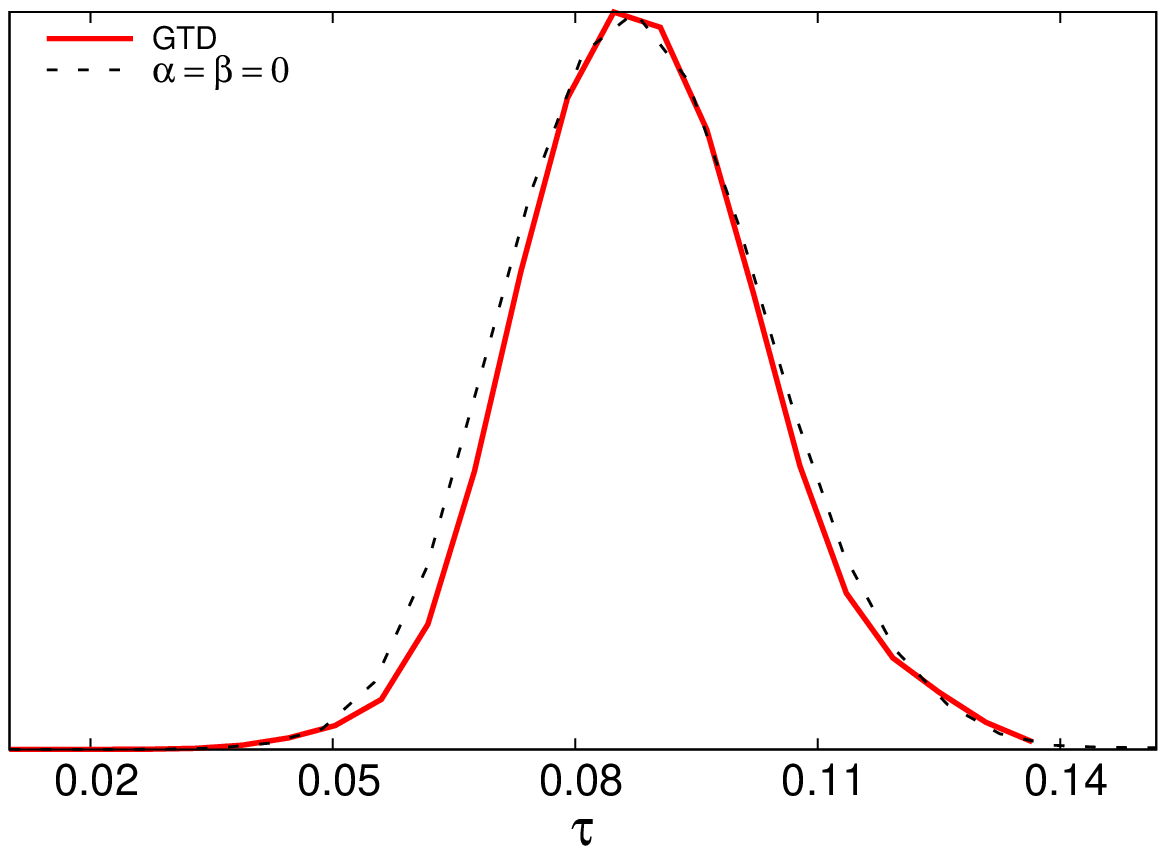}
\includegraphics[width=1.9in]{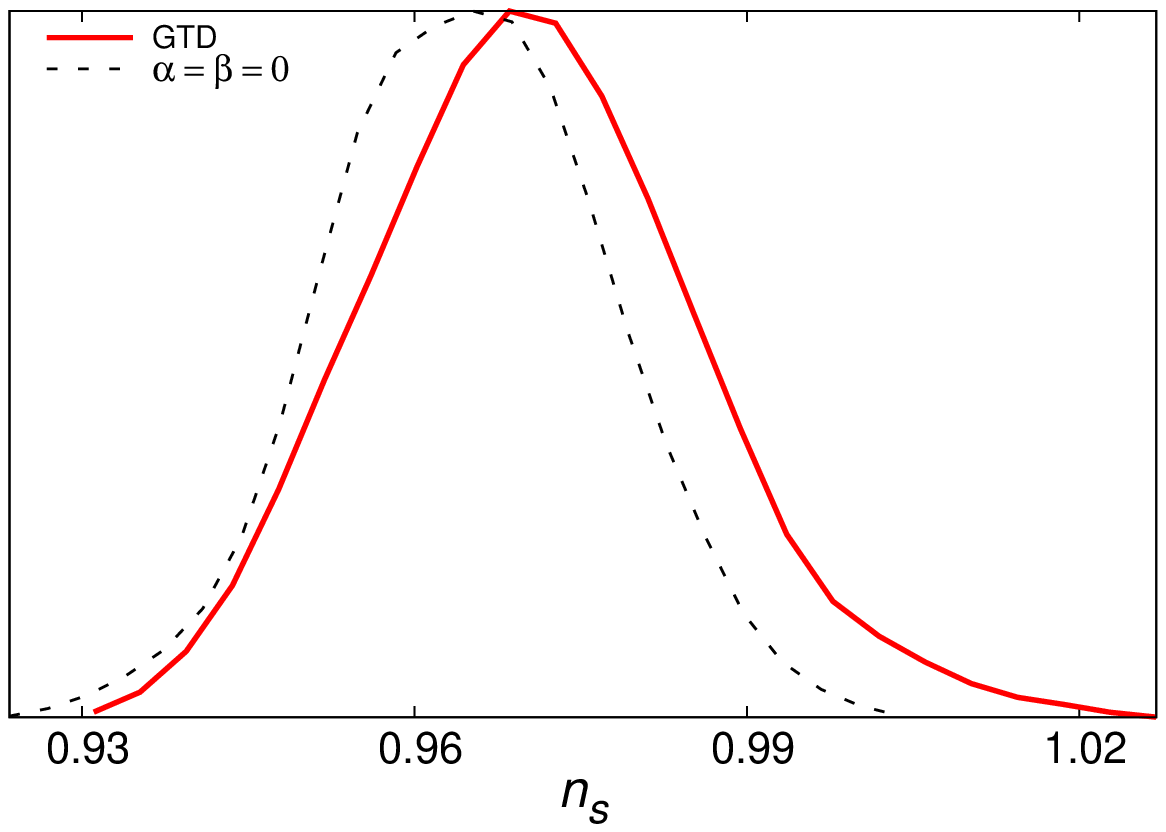}
\includegraphics[width=1.9in]{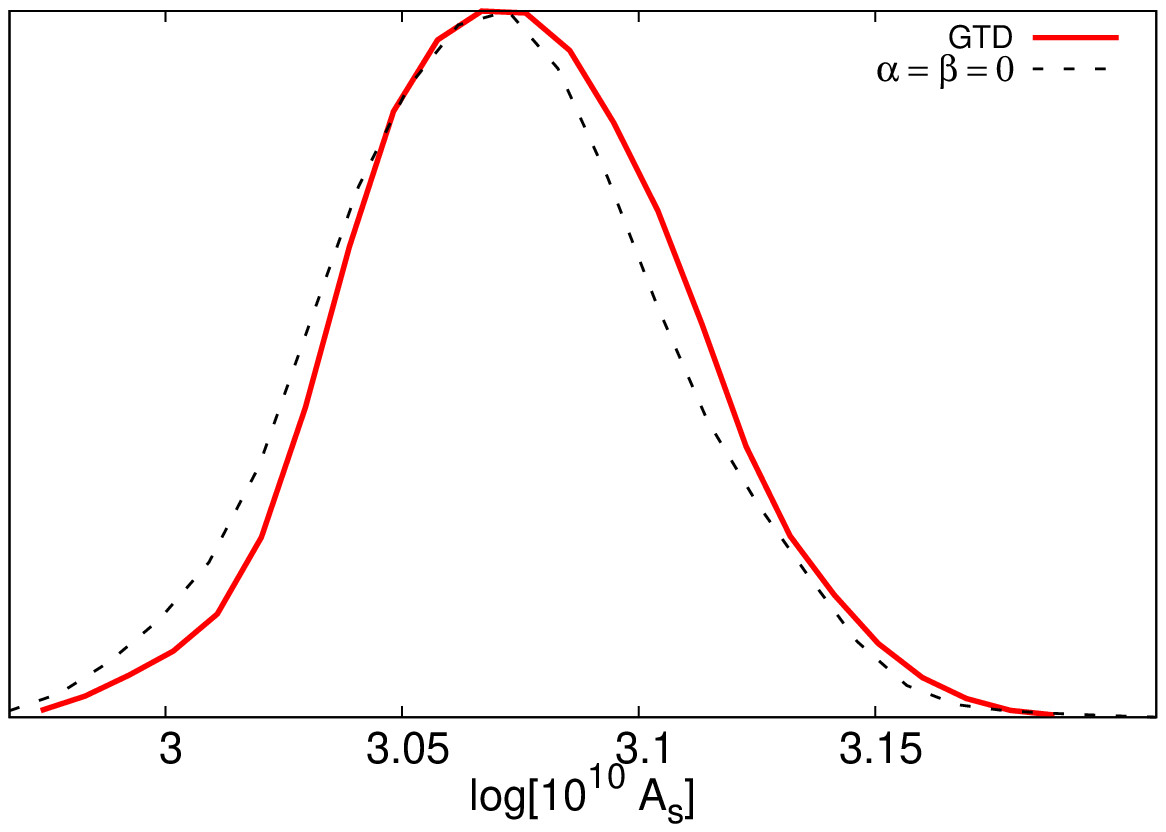}
\includegraphics[width=1.9in]{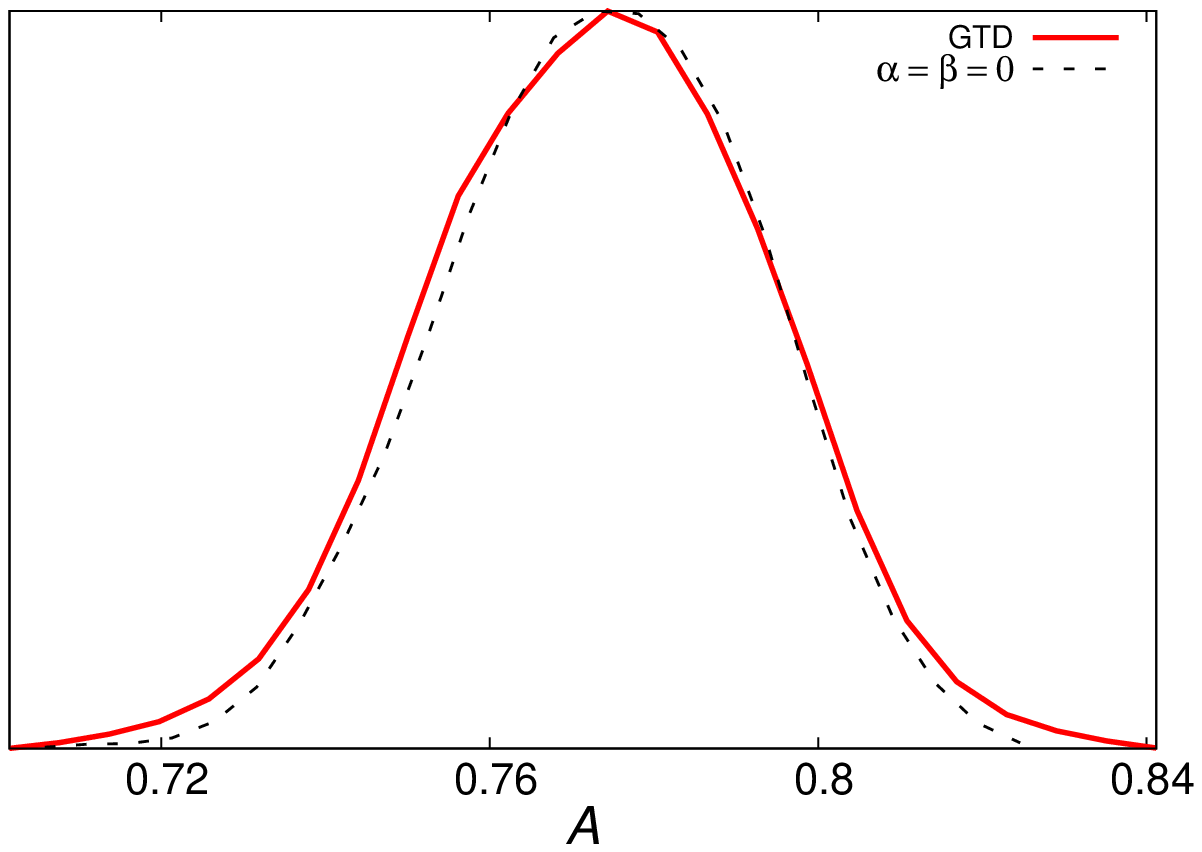}
\includegraphics[width=1.9in]{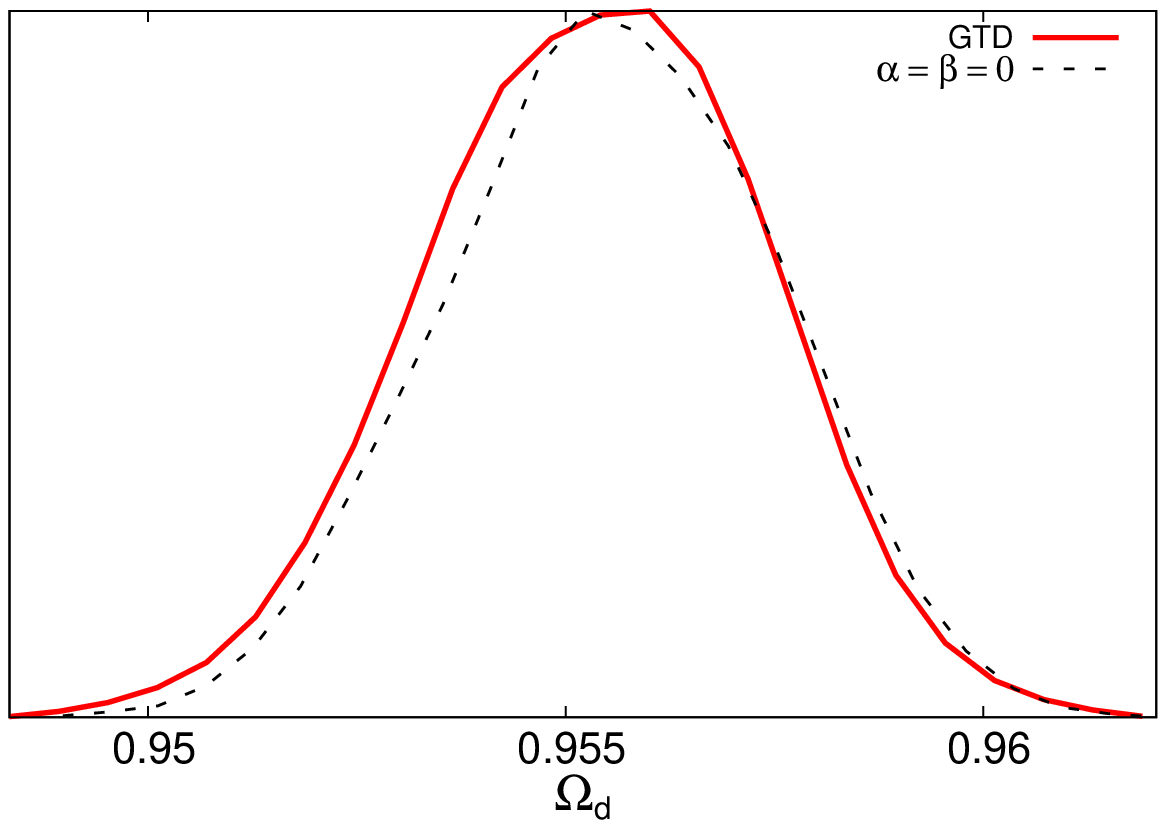}
\includegraphics[width=1.9in]{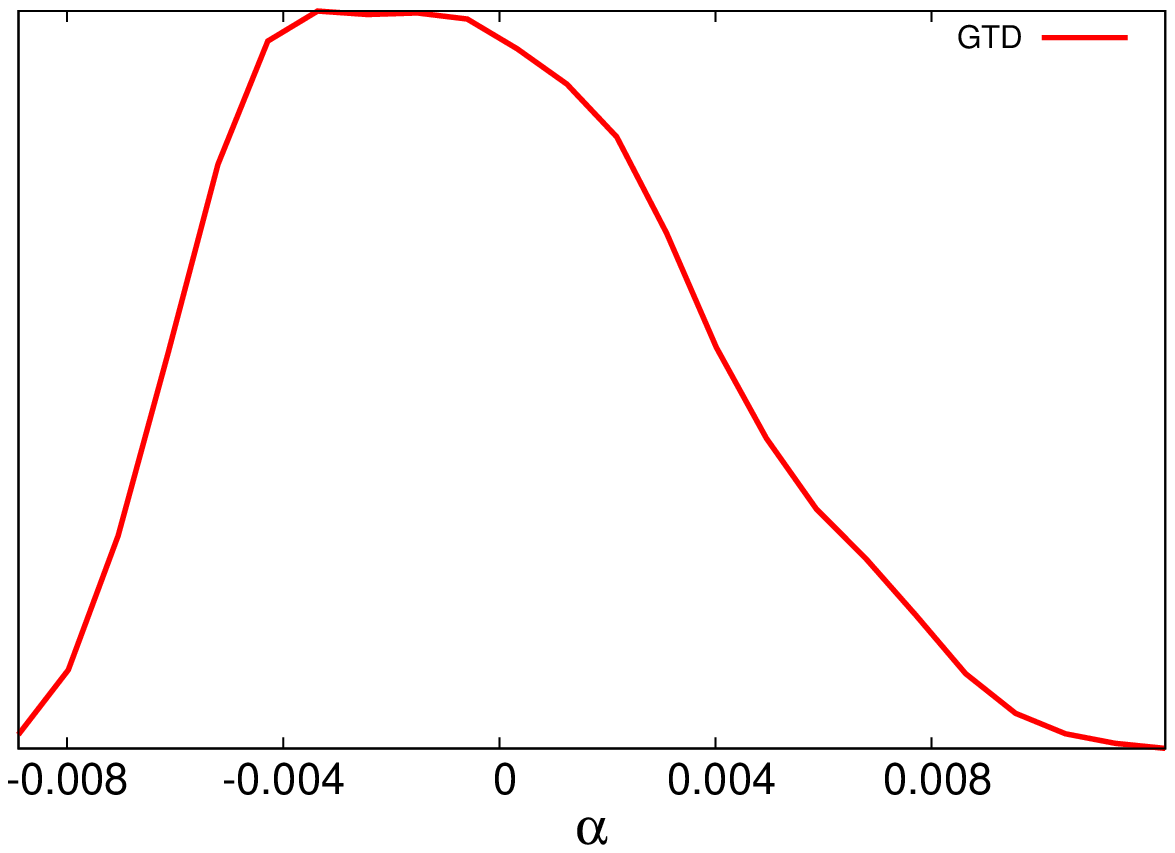}
\includegraphics[width=1.9in]{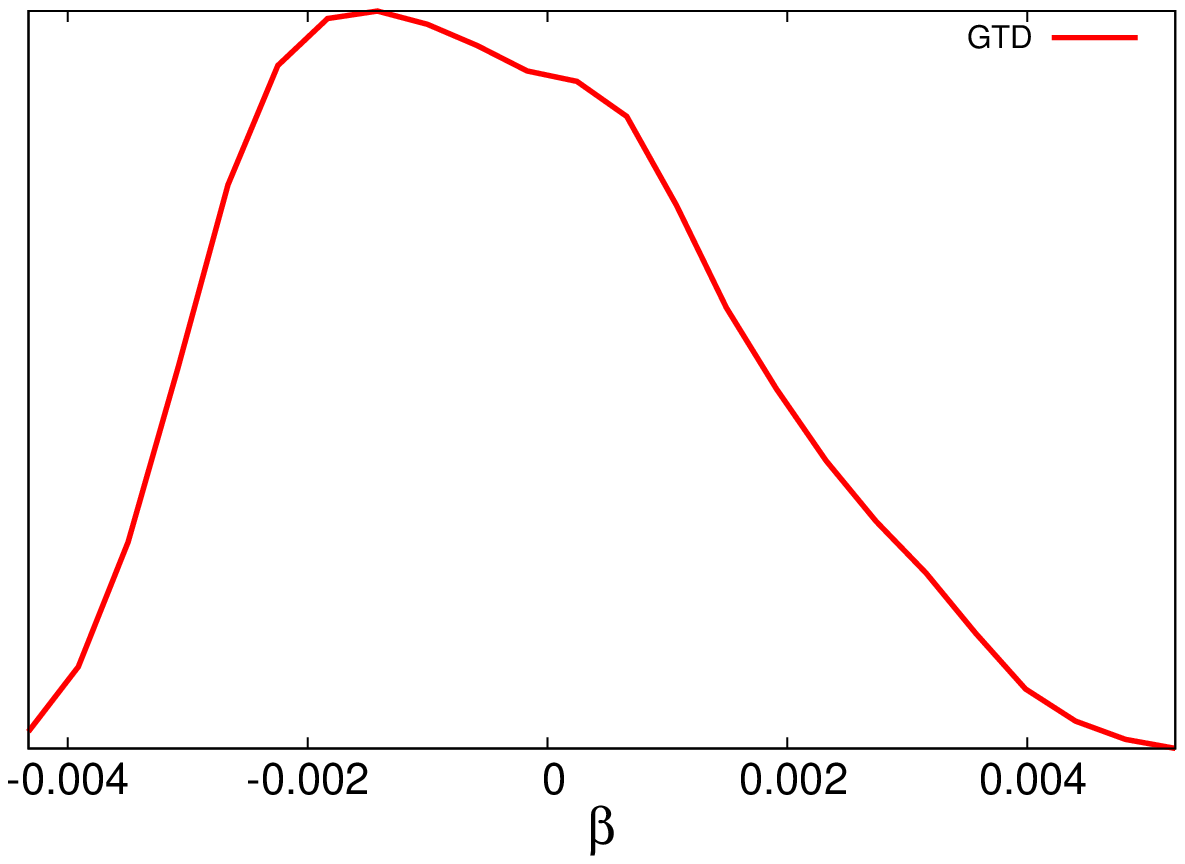}
\caption{1-dimensional marginalized probability for the complete set of parameters ex\-plo\-red with MCMC and the derived parameter
$\Omega_d$. The data used are the WMAP seven-year results, Union 2 data set supernovae compilation and a prior of HST on the Hubble constant. 
The dashed curves are obtained imposing $\alpha = \beta = 0$.
\label{fig:1dpdf}
}
\end{center} 
\end{figure}

In figure \ref{fig:1dpdf} 
the 1-dimensional posteriors of the explored space parameter $\mathcal{M}$ and the derived parameter $\Omega_d$ are shown. For comparison, 
the results for the dark fluid model are also plotted. To translate the latter quantities to the $\Lambda$CDM model language, one only needs 
to use the equations $\mathcal{A} = 1/(1+ \Omega_{DM}/\Omega_{\Lambda})$ and $\Omega_d = \Omega_{DM} + \Omega_{\Lambda}$, for details see 
\cite{Aviles:2011ak}. In table 1 we present the summary of the results at 0.68 confidence level (c.l.). 

We obtain that the free parameters of the GTD unified fluid have to take values of the order of $10^{-3}$ or 
lesser, although in principle they could be as large as causality allows (for the cases $\alpha= \beta$, this is $\alpha < 1$). This constraints are 
in agreement with those found for the generalized Chaplygin gas in the literature; see e.g. \cite{Piattella:2009da, Gorini:2007ta}.

\begin{table*}
\begin{center}
\caption{{\small Summary of constraints. The upper panel contains the parameter spaces ex\-plo\-red with MCMC. 
The bottom panel contains derived parameters. The data used are the WMAP seven-year data, Union 2 compilation and HST.}}          
\begin{tabular}{c|c} 
\hline\hline 
{\small Parameter}   &   Best fit${}^a$ \\ [1.5ex]
\hline 
\hline
 {\small $10^2 \Omega_b h^2$}         & $\qquad$ {\small $2.231$}{\tiny${}_{-0.123}^{+0.163}$} $\qquad$                
                                                        \\[0.8ex] 
{\small $\theta$}                     & {\small $1.038$}{\tiny${}_{-0.005}^{+0.007}$}                
                                        \\[0.8ex]
{\small $\tau$}                       & $\quad${\small$0.0892$}{\tiny${}_{-0.0036}^{+0.0038}$}$\quad$      
                                         \\[0.8ex]
{\small $10^3 \alpha$}                & {\small $-3.21$}{\tiny${}_{-5.00}^{+11.92}$}          
                                        \\ [0.8ex] 
{\small $10^3 \beta$}                 & {\small $-1.56$}{\tiny ${}_{-2.38}^{+5.59}$}                                
                                        \\[0.8ex]
{\small $\mathcal{A}$}                & {\small $0.768$}{\tiny ${}_{-0.047}^{+0.051}$}                                
                                        \\[0.8ex]
{\small $n_s$}                        & {\small $0.963$}{\tiny ${}_{-0.027}^{+0.047}$}    
                                       \\[0.8ex]
$\quad$ {\small $\log[10^{10} A_s]$}$\quad$   & {\small $3.075$}{\tiny ${}_{-0.085}^{+0.086}$}        
                                        \\[0.8ex]
\hline 
{\small $\Omega_d$}          & {\small $0.955$}{\tiny ${}_{-0.004}^{+0.005}$}               
                              \\ [0.8ex] 
{\small $t_0\,{}^b$}         & {\small $13.84$}{\tiny ${}_{-0.35}^{+0.23}\,$}  
                               \\ [0.8ex]
{\small $H_0\,{}^c$}         & {\small $70.41$}{\tiny${}_{-3.79}^{+5.84}$}                 
                               \\ [0.8ex]
\hline\hline
\end{tabular}

{\small 

a. The maximum likelihood of the sample. The quoted errors show the $0.68$ c.l.
 
b. The Age of the Universe ($t_0$) is given in gigayears.

c. $H_0$ is given in Km/s/Mpc.

}
\label{table:nonlin} 
\end{center}
\end{table*}

There is a non-linear effect that we have not considered so far and that arises from the fact that in general the relation $\langle P \rangle = P(\langle \rho \rangle, a)$ 
does not hold. Therefore, when some scale grows and becomes non-linear, the naive averaging procedure is no longer valid.
 This effect has been investigated in the past in \cite{Avelino:2003ig,Beca:2007rd}; for alternative approaches see \cite{Sussman:2008wx, Roy:2009cj}. 
In fact, in our case it follows a relation
\begin{equation}
\langle P \rangle = - \mathcal{C} a^{-3(\alpha-\beta)}\langle \rho \rangle^{-\alpha} (1-\alpha \delta + \mathcal{O}_2(\delta)),
\end{equation}
between averages quantities. It is clear that considering these effects complicates the calculations considerably and it is out of the scope of this work to treat 
them accurately. We expect the free parameters of the GTD fluid to be even more constrained  by the corrections induced by this non-linear effect. 


\section{Concluding remarks} 
\label{sec:con}

In this work, we applied the formalism of GTD to construct models of fluids that can be used as 
gravitational sources in the context of relativistic cosmology. First, we considered the simplest 
GTD fluid that corresponds to a thermodynamic system whose equilibrium manifold is flat, and found that 
it can be used to generate all the fluids of the standard cosmological model. We also discussed the possibility 
of considering the dark energy fluid as a non-interacting thermodynamic system with negative heat capacity and non-extensive thermodynamic variables.

We then investigated a GTD fluid whose thermodynamic curvature is non-zero in general, indicating the presence of internal thermodynamic interaction. 
It turned out that this fluid leads to a new cosmological model whose equation of state contains as special cases the generalized Chaplygin gas, 
the dark fluid model, and the polytropic fluids. We showed that it is possible to  interpret this new GTD fluid as corresponding to a unified model 
for dark matter and dark energy. To prove this, we used the Friedmann equations to perform a detailed analysis of the behavior of the state parameter 
of the GTD dark sector and of the corresponding scale factor. The obtained results are in accordance with current cosmological observations. The main 
difference between the generalized Chaplygin gas and the GTD fluid consists in the behavior of the adiabatic speed of sound. Although at the cosmological 
background level this difference does not lead to any considerable consequences, the perturbation of the background cosmology shows an essential 
difference at the level of the density contrasts. The square of the adiabatic speed of sound is always positive for the Chaplygin gas model 
but negative in general for the GTD fluid, leading 
to an enhancement of the structure growth in the latter case. Moreover, the analysis of the CMB angular power spectrum shows that deviations from the 
$\Lambda$CDM model appear only at large scales. Finally, we find the best fit parameters of the GTD fluid by 
using current observational data and show that the parameters $\alpha$ and $\beta$ must be of the order of $10^{-3}$ or lesser.

We conclude that from GTD it is possible to obtain fundamental equations for thermodynamic systems that can be used to develop physically reasonable 
cosmological models. In this work, we analyzed only two simple GTD fluids. It would be interesting to study more complicated GTD solutions and their 
interpretation in the framework of relativistic cosmology. 

The  microscopic nature of the GTD dark fluid is unknown, as much as the dark matter and dark energy in the $\Lambda$CDM model. In this work we have focused on its geometrical
description by using the formalism of GTD.
\section{Acknowledgments}

We would like to thank J.L. Cervantes-Cota, C.S. Lopez-Monsalvo, F. Nettel, A. S\'anchez, and R. Sussman for helpful comments and discussions. 
This work was supported in part by DGAPA-UNAM, grant No. IN166110, and Conacyt-Mexico, grant No. 166391.

\bibliographystyle{JHEP}
\bibliography{bibsgtd.bib}

\end{document}